\documentclass[prd,aps,twocolumn,preprintnumbers, showpacs, nofootinbib,notitlepage]{revtex4-1}
\usepackage{amssymb,amsthm,amsmath}
\usepackage{textcomp}
\usepackage{slashed}    
\usepackage{verbatim}
\usepackage[normalem]{ulem} 
\usepackage{hyperref}
\usepackage{soul}

\usepackage{empheq} 
\usepackage{rotating}   
\usepackage{multirow}   

\usepackage{subcaption}
\usepackage{epsfig}
\begin{document}
	
	\raggedbottom
	
	\title{ 
		Strange   Quark Electric Dipole Moment  with Topological Anomalies  }

	\author{   Chao-Qiang Geng}	
	\affiliation{School of Fundamental Physics and Mathematical Sciences, Hangzhou Institute for Advanced Study, UCAS, Hangzhou 310024, China}
	\author{   Xiang-Nan Jin } 
	\affiliation{School of Fundamental Physics and Mathematical Sciences, Hangzhou Institute for Advanced Study, UCAS, Hangzhou 310024, China}
	\author{   Chia-Wei Liu  }
	\affiliation{School of Fundamental Physics and Mathematical Sciences, Hangzhou Institute for Advanced Study, UCAS, Hangzhou 310024, China}
	
	\author{ Bin  Wu  }
	\affiliation{School of Fundamental Physics and Mathematical Sciences, Hangzhou Institute for Advanced Study, UCAS, Hangzhou 310024, China} 
	
	\date{\today}
	
	\begin{abstract}
We introduce an anomaly-based framework to probe quark electric dipole moments in exclusive hadronic final states produced in $e^+e^-$ annihilation.
Chern--Simons--induced anomalous couplings yield a clean T-odd asymmetry $A_T$ from interference between the Standard-Model amplitude and dipole-moment contributions. 
As an application, $\gamma^\ast\to K^+K^-\pi^0$ provides direct sensitivity to the strange-quark EDM $d_s$ in the chiral limit.
The precision on $d_s$ can reach $\mathcal{O}(10^{-16})\,e\cdot\mathrm{cm}$ with existing CMD--3 data and $\mathcal{O}(10^{-18})\,e\cdot\mathrm{cm}$ using current $J/\psi$ samples at BESIII, improving the current direct constraint from hyperon EDM by two to three orders. 
	\end{abstract}
	\maketitle

	\section{
		Introduction
	}
	
Electric dipole moments (EDMs) of fermions are powerful null tests of CP violation~\cite{Smith:1957ht,Bernreuther:1990jx}.
In particular, nucleon EDM searches provide the most stringent constraints on CP violation in QCD, implying
$|\overline{\theta}|\lesssim 10^{-10}$ for the QCD vacuum angle~\cite{He:1989mbz,Abel:2020pzs,Chupp:2017rkp}.
Complementary bounds on hadronic CP violation also arise from atomic EDM searches~\cite{Lamoreaux:1987zz}.
The electron EDM currently provides the tightest constraint on many classes of physics beyond the Standard Model (SM)~\cite{Roussy:2022cmp,Barr:1990vd,Ramsey-Musolf:2006evg,Li:2010ax,Jung:2013hka,Inoue:2014nva,Shu:2013uua,Li:2021xmw,He:1992dc,Dorsner:2016wpm,Fuyuto:2018scm,Dekens:2018bci}.

Nevertheless, EDMs of quarks beyond the first generation remain subject to little direct experimental constraint~\cite{Huang:2025ghw,Belle:2021ybo,Vecchi:2025jbb}. 
In the lepton sector, the muon EDM provides a direct probe of CP violation beyond the first generation  from storage-ring measurements~\cite{Muong-2:2008ebm}. 
Very recently, the BESIII collaboration reported the first direct bound on the \(\Lambda\) hyperon EDM from 
\(J/\psi \to \Lambda \overline{\Lambda}\)~\cite{BESIII:2025vxm}:
\begin{equation}\label{eq1}
	\left| d_\Lambda(m_{J/\psi}^2) \right|
	< 6.5 \times 10^{-19}\, e\cdot\mathrm{cm}\,,
\end{equation}
where $\sqrt s $ is the virtual-photon invariant mass. 
Because QCD is nonperturbative in this kinematic regime, a direct relation between \(d_\Lambda(s)\) and the strange-quark EDM \(d_s\) 
can only be established at the zero-recoil point with an on-shell photon;  \(d_\Lambda(0)=d_s\)~\cite{He:1989mbz} in the quark model. 
Once the momentum dependence of \(d_\Lambda(s)\) is taken into account, the resulting constraint on \(d_s\) becomes weaker, since
$ 
d_\Lambda(m_{J/\psi}^2)=5.29\times10^{-4}\, d_s\, e\cdot \mathrm{cm}
$~\cite{Chen:2025rab}. 
On the other hand, the neutron EDM provides a significantly stronger indirect constraint~\cite{Gupta:2018lvp,Dragos:2019oxn}:
\begin{eqnarray}
	d_n &=&  
	-(1.5 \pm 0.7) \times 10 ^{-16} \overline{\theta }\, e\mathrm{\cdot cm} -  
	( 0.20\pm 0.01 )d_u
	\nonumber\\ 
	&& 
	+ ( 0.78 \pm 0.03 ) d_d  + ( 2.7 \pm 1.6 )\times 10^{-3} d_s \,,
\end{eqnarray}
with the experimental bound
$|d_n| < 1.8 \times 10 ^{-26}e\cdot\mathrm{cm}$~\cite{Abel:2020pzs}. 
This relation makes it clear that $d_n$ alone cannot disentangle the individual quark EDM contributions. In particular, cancellations among different terms may substantially weaken the direct sensitivity to $d_s$, thereby leaving room for the strange-quark EDM to be probed more effectively in complementary observables. 

\begin{figure}
	\centering
	\includegraphics[width=0.49\linewidth]{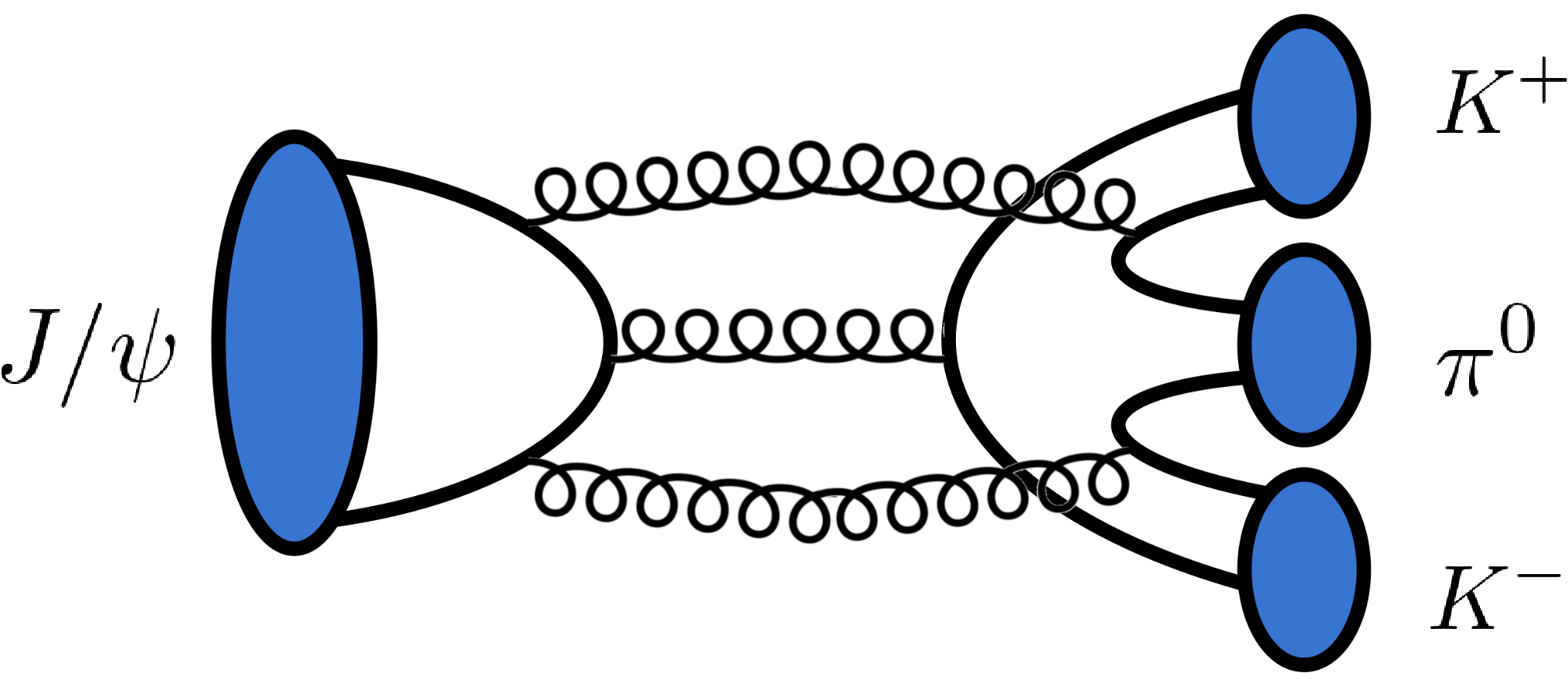}
	\includegraphics[width=0.49\linewidth]{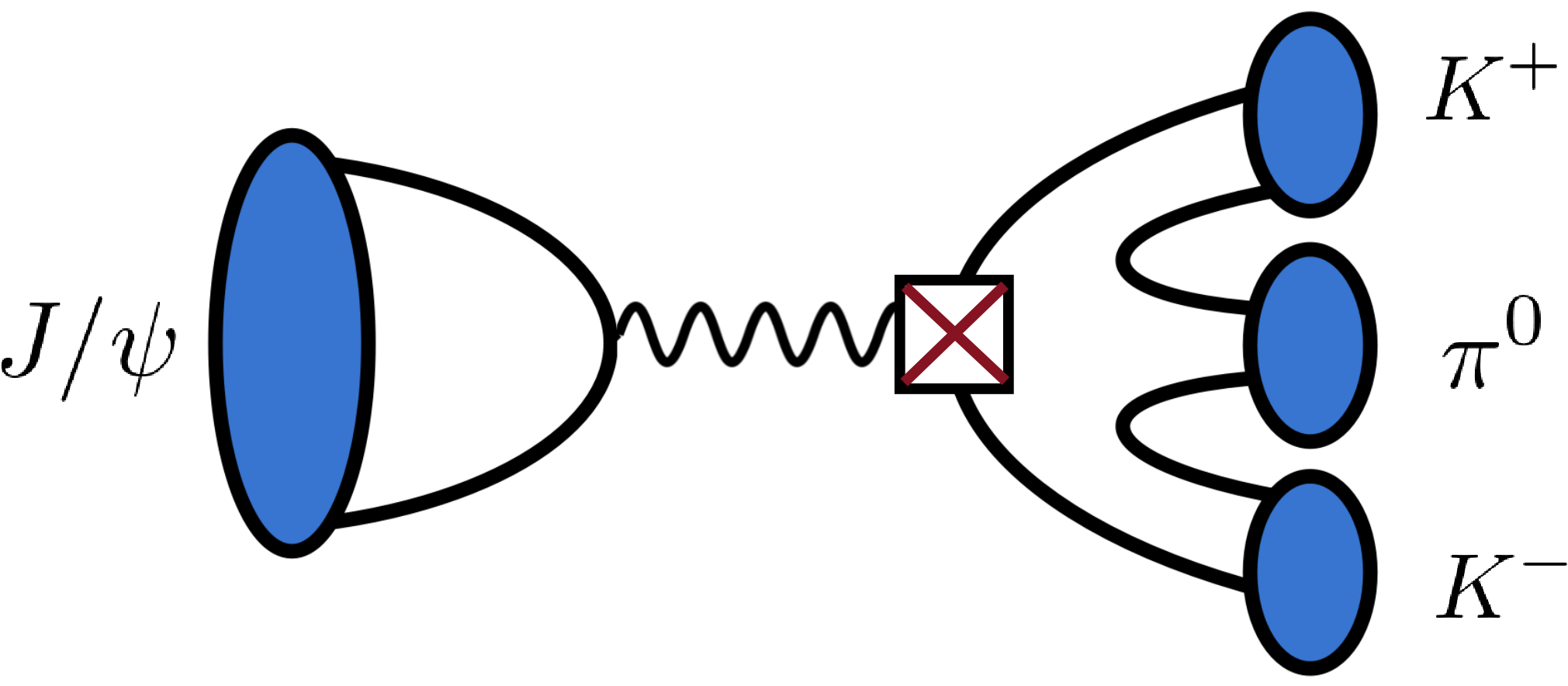}
\caption{Schematic structure of the amplitudes for $J/\psi \to \gamma^\ast\!\to K^+K^-\pi^0$.
	The left panel shows the SM contribution mediated by three-gluon annihilation.
	The right panel shows the EDM-induced contribution, where the boxed cross denotes the strange-quark EDM. } 
	\label{fig:edm}
\end{figure}

In this work we propose a new way to probe quark EDMs using CP-odd angular correlations in exclusive hadronic final states produced in $e^+e^-$ annihilation.
In particular, $J/\psi$  may appear as the intermediate state as shown in FIG.~\ref{fig:edm}.
The key theoretical handle is the Wess--Zumino--Witten (WZW) term~\cite{Wess:1971yu,Witten:1983tw} in chiral perturbation theory ($\chi$PT)~\cite{Gasser:1983yg,Gasser:1984gg,Leutwyler:1993iq,Weinberg:1978kz}, which encodes the QCD chiral anomaly and fixes the structure and normalization of odd-intrinsic-parity amplitudes~\cite{Kaymakcalan:1983qq,Fujiwara:1984mp,Chou:1983qy,Wu:1986pr}.
This anomaly protection provides a controlled SM baseline against which dipole-induced CP violation can be isolated.
As an application, we focus on the strange-quark EDM $d_s$ in $e^+e^-\to K^+K^-\pi^0$, where the low-energy $\gamma^\ast\to K^+K^-\pi^0$ amplitude is governed by the WZW interaction and remains fixed in the chiral limit even with resonances included~\cite{Dax:2020dzg,Harada:2003jx,Geng:2025fuc,Geng:2026rbz}.
Compared with hyperon channels, this mode has a larger branching fraction and avoids the loss of efficiency from subsequent decays,
which would otherwise reduce the effective statistics by $(\alpha_{\mathrm{sub}}{\cal B}_{\mathrm{sub}})^2$~\cite{Du:2024jfc,Fu:2023ose},
where ${\cal B}_{\mathrm{sub}}$ and $\alpha_{\mathrm{sub}}$ denote the branching fraction and decay parameter of the subdecays.

This paper is organized as follows. 
In Sec.~\ref{sec2}, we construct the CP-odd observable. 
In Sec.~\ref{sec3}, we match \(d_s\) onto \(\chi\)PT and present the WZW prediction. 
In Sec.~\ref{sec4}, we estimate the attainable experimental precision. 
We conclude in Sec.~\ref{sec5}.

	\section{
		CP-odd observables  
	}\label{sec2}
As a concrete example, 
consider $\gamma^*(q)\to K^+(p_+)\,K^-(p_-)\,\pi^0(p_0)$. 
	We employ the symmetric set of kinetic variables 
	$ 
	s = q^2,
	$ and $s_{\pm}
	=  (p_\pm+p_0)^2.
	$
	The total amplitude is parameterized as 
	\begin{equation}
		\label{eq:amp-total}
		{\cal M}(\gamma^*\to K^+K^-\pi^0)
		=
		\epsilon_\mu(q)\,J^\mu ,
	\end{equation}
	where $\epsilon$ is the polarization vector of the virtual photon.
	Using a gauge-invariant decomposition, the hadronic current is parameterized as
	\begin{equation}
		\label{eq:J-decomp}
	 	J^\mu
	= 
		F_V \Big(s, s_+ , s_- \Big)  V^\mu\!
		+\!
		F_A\!\Big(s, s_+ , s_-   \Big)  A^\mu , 
	\end{equation}
	with the  vector and axial vectors are 
	\begin{eqnarray}
		\label{eq:AV-basis}
		V^\mu
		&=&
				\,\epsilon^{\mu\nu\alpha\beta}\,
		p_{+\nu}\,p_{-\alpha}\,p_{0\beta},
\nonumber\\
		A^\mu
		&=&
		(q\!\cdot\! p_0)\,(p_+-p_-)^\mu
-
\big(q\!\cdot\!(p_+-p_-)\big)\,p_0^\mu, 
	\end{eqnarray}
respectively, 
 satisfying $q_\mu V^\mu=q_\mu A^\mu=0$.
	Invariance under C-parity demands $F_V$ and $F_A$ to be invariant under exchanging $s_+$ and $s_-$; see the figures in Ref.~\cite{BESIII:2019apb}.    Since the final state has negative intrinsic parity, one sees that $F_V$ and $F_A$ are CP-even and CP-odd, respectively.  
	
	In the $\gamma^*$ centre-of-mass frame, $q^\mu=(\sqrt{s},\vec 0)$ and
	$p_i^\mu=(E_i,\vec p_i)$. 
	The amplitude can be written as
	\begin{equation}
		{\cal M}=  - \vec\epsilon\cdot\Big(
		F_V  \vec V
		+
		F_A \vec A
		\Big),  
	\end{equation}
	with 
	$	\vec A   = 
	\sqrt{s}\,\big[
	E_0\,(\vec p_+-\vec p_-)
	+ (E_+-E_-)\,\big(
	\vec p _+ + \vec p _- 
	\big)\big]
	$ 
	and 
	$
	\vec V =  \sqrt{s}\,(\vec p_-\times \vec p_+)
	$. 
	Taking the unpolarized $e^+ e^- \to \gamma^*$ case, we sum over the two transverse photon polarizations and 
	obtain
	\begin{equation} 
		\overline{|{\cal M}|^2} = 
		\frac12\sum_{\lambda=\pm}|{\cal M}|^2
		=
		\frac12\Big|\big(F_V\,\vec V +  F_A\,\vec A\big)_\perp \Big|^2,
	\end{equation}
	where $\vec X_\perp \equiv \vec X-(\vec X\cdot\hat z)\hat z$ denotes the component transverse
	to the beam axis $\hat z$.   
	The interference term proportional to
	$\Re(F_V F_A^*)\,\vec V_\perp  \cdot \vec A_\perp  $
	is C-even and CP-odd, and will be shown to be induced by $d_s$. 
	
\begin{figure}
		\centering
		\includegraphics[width=0.9\linewidth]{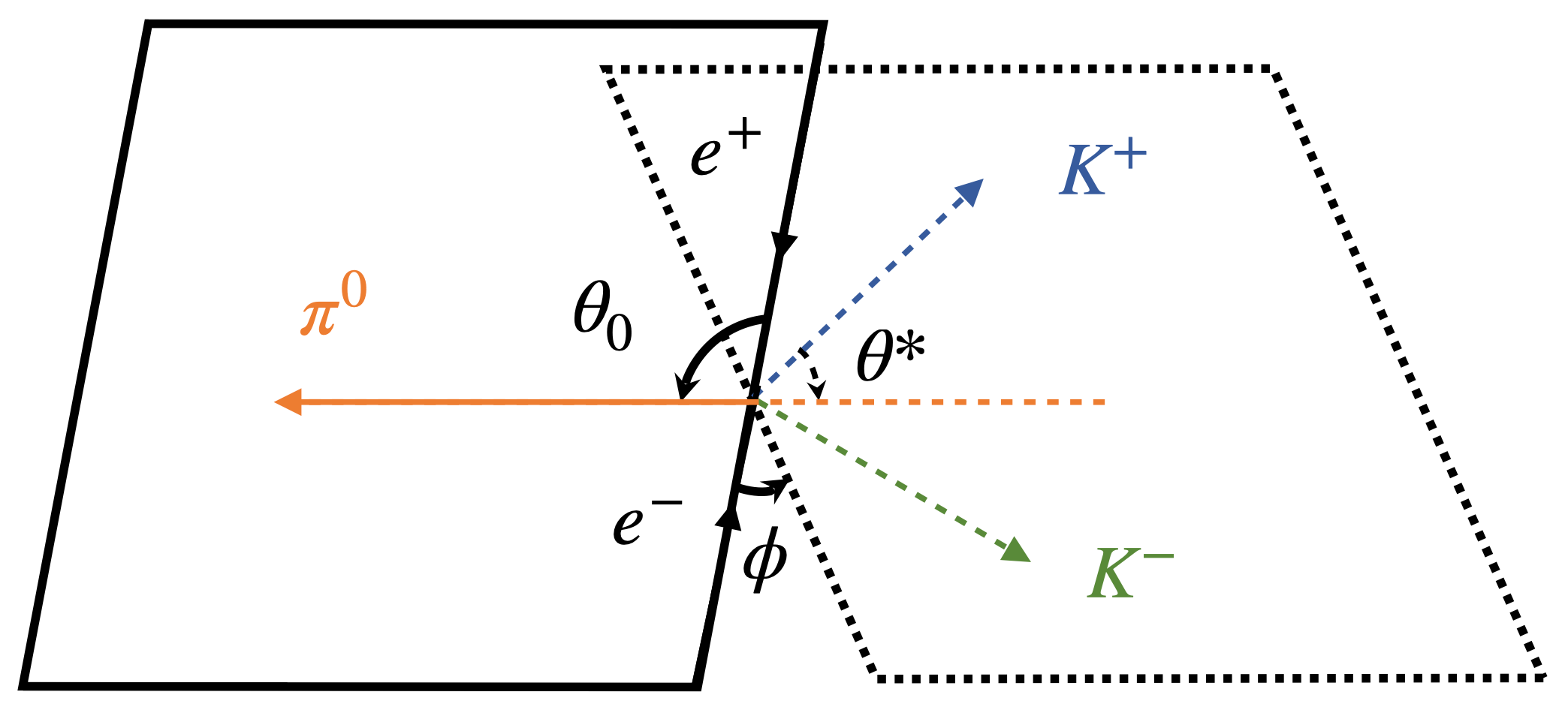}
		\caption{The definitions of the angles in Eq.~\eqref{angles}.
			The angle $\theta^0$ is the angle between the momenta of $e^+$ and $\pi^0$ in the center-of-mass frame.
			The angle $\theta^\ast$ is the angle between $\pi^0$ and $K^+$ in the $K^+K^-$ rest frame.
			The angle $\phi$ is the dihedral angle between the two planes spanned by $(\pi^0,e^+,e^-)$ and $(\pi^0,K^+,K^-)$. 
	}
		\label{fig:angles}
	\end{figure}

	The fully differential decay distribution  in the $\gamma^*$ rest frame is  given by 
	\begin{equation}\label{angles}
		\frac{d\Gamma}{  d\Phi  }
		=
		\frac{1}{(2\pi)^4}\,
		\frac{|\vec p_0|\,|\vec p_+^{\,*}|}{16\,s}\;
		\overline{|{\cal M}|^2},
	\end{equation} 
	where
	$ 
	d \Phi \equiv dm_{KK} \, d\cos \theta _ 0\, d\Omega_+^{*}.
	$
	Here,
	$\theta_0$ is the inclined angle of the three-momentum of \(\pi^0\) 
	and $\gamma^*$ polarization 
	in the \(\gamma^\ast\) rest frame, while
	\(\Omega_+^\ast=(\theta^\ast,\phi)\) and \(\vec p_+^{\,\ast}\) are those of \(K^+\) in the \(K^+K^-\) rest frame; see FIG.~\ref{fig:angles}. 
	
	To isolate the contribution linear in $F_A$ without assuming a specific kinematic
	dependence, we define the signed interference observable
	\(
	\vec V_\perp \cdot\vec A_\perp 
	\)
	and construct the counting asymmetry
	\begin{equation}
		\label{eq:AT-WD}
		A_T \equiv
		\frac{
			N\big(   \vec V_\perp  \!\cdot\!\vec A_\perp  >0\big)-N\big( \vec V_\perp  \!\cdot\!\vec A_\perp  <0\big)
		}{
			N\big(   \vec V_\perp \!\cdot\!\vec A_\perp  >0\big)+N\big(\vec V_\perp \!\cdot\!\vec A_\perp  <0\big)
		}\,.
	\end{equation}
	Only 
	the interference term
	$\Re(F_V F_A^*)\,(\vec V_\perp  \cdot \vec A_\perp )$ contributes, providing a robust probe of $\Re(F_V F_A^*)$.  We note that $A_T$ is C-even and P-odd, and $\gamma^*\to K^+K^-\pi^0$ is self-conjugate under charge conjugation; hence $A_T$ is a genuinely CP-odd observable. In particular, final-state interactions (with or without  weak interactions at the leading order) cannot induce a nonzero $A_T$. 

	Finally, we note that 
	the framework here can be straightforwardly extended to the one with $J/\psi $ as the initial state, as its $J^{PC}$ is identical to $\gamma^*$.

	\section{
		Mapping $d_s$ to 	$\chi$PT
	}\label{sec3}
	
We start from the flavor-conserving CP-violating strange-quark dipole operator 
\begin{equation}\label{quarkLevel}
	\mathcal{L}_{\text{eff}}
	=
	\frac{d_s}{2}\,
	\tilde F^{\mu\nu}\,
	\bar s\,\sigma_{\mu\nu}\, s ,
\end{equation}
where \(d_s\) parametrizes the strange-quark electric dipole operator induced by new physics.
Although the $d_n$  bound yields a much tighter indirect constraint on \(d_s\), it is tied essentially to the \(q^2=0\) limit. The process considered here instead probes the timelike region, where the relation between the measured hadronic observable and the underlying strange-quark EDM is intrinsically \(q^2\)-dependent. An observable effect in this channel would therefore provide genuinely new information, while at the same time testing whether nontrivial cancellations among CP-odd contributions to $d_n$~\cite{Bhattacharya:2015esa}. 

The QCD Lagrangian enjoys a global \(SU(3)_L \otimes SU(3)_R\) flavor symmetry, with
\(q_{L}\to g_{L}\,q_{L}\), \(q_{R}\to g_{R}\,q_{R}\)  and \(q=(u,d,s)^T\).
To make chiral symmetry manifest in the presence of \(d_s\), we introduce spurion fields
\(t_\pm\) that transforming as 
$
	t  \to g_L\, t \, g_R^\dagger, 
$
With these spurions, the effective Lagrangian can be rewritten as
\begin{equation}
	\mathcal{L}_{\text{eff}}
	=
	\frac{d_s}{2}\,
	\tilde F^{\mu\nu}\,
	\Big(
	\bar q_L\,\sigma_{\mu\nu}\, t \, q_R
	+
	\bar q_R\,\sigma_{\mu\nu}\, t ^\dagger  \, q_L
	\Big).
\end{equation}
Matching to Eq.~\eqref{quarkLevel}, one sets
\(
t  =\mathrm{diag}(0,0,1)
\)
in flavor space.  With the spurion transformation rules,
\(\mathcal{L}_{\text{eff}}\) is formally invariant under \(SU(3)_L\otimes SU(3)_R\).

	To match onto chiral fields, we note that $ U = \exp ( 2 i P  / f_P )$  with 
	\begin{eqnarray}
		P  \!=\! \!\! \  \left[ \begin{array}{ccc}
			\frac{\eta_0}{\sqrt{3}} + \frac{\eta_8}{\sqrt{6}} + \frac{\pi^0}{\sqrt{2}} & \pi^{+} & K^{+} \\
			\pi^{-} & \frac{\eta_0}{\sqrt{3}} + \frac{\eta_8}{\sqrt{6}} - \frac{\pi^0}{\sqrt{2}} & K^0 \\
			K^{-} & \bar{K}^0 & \frac{\eta_0}{\sqrt{3}} - \frac{2 \eta_8}{\sqrt{6}}
		\end{array} \right]  \! , 
	\end{eqnarray}
	which transforms as 
	$
	U \to g_L\,U\,g_R^\dagger . 
	$
	The lowest-order chiral realization of the strange-quark tensor currents is 
	\begin{eqnarray}
		{\cal L} 
		&=&   
		\frac{ d_s }{2} 
		\tilde 	F^{\mu\nu}  
		\frac{    
			\Lambda_2 } { 8}
		T_{\mu \nu} (t )
		\nonumber\\	
		T_{\mu \nu} (t )  &=& 	 i  	  
		\mathrm{Tr}\!\Big[(t U^\dagger+Ut ^\dagger )\,[\alpha_\mu,\alpha_\nu]\Big]
		\\ &  &\,    - \,\epsilon_{\mu\nu\lambda\sigma}\,
		\mathrm{Tr}\!\Big[(t U^\dagger- Ut ^\dagger )\,\alpha^\lambda \alpha^\sigma \Big]
		\,,  \nonumber 
	\end{eqnarray}
	where 
	$ 	\alpha_\mu=-i\,\partial_\mu U\,U^\dagger$
	and $\Lambda _{2}$ is a low-energy constant.   
	The tensor  
	$T_{\mu\nu} $  is constructed to satisfy
	\(\tfrac12\,\epsilon^{\mu\nu\alpha\beta}T_{\alpha\beta}(t )=  \,  \,T^{\mu\nu}( - i t )\),
	mirroring \(\tfrac12\,\epsilon^{\mu\nu\alpha\beta}\sigma_{\alpha\beta}=-\,i\,\sigma^{\mu\nu}\gamma_5\)~\cite{Cata:2007ns}.

	Ideally, one searches for channels in which the chiral approximation is reliable and all incoming and outgoing states are on shell.  Imposing flavor conservation, the simplest such radiative channel is $\eta^{(\prime)}\to\pi^+\pi^-\gamma$~\cite{Geng:2002ua,Gan:2020aco,Sanchez-Puertas:2018tnp} at leading order.  Nevertheless, $C$-parity symmetry forces the $\pi^+\pi^-$ system to be an isovector, and hence $d_s$ cannot induce this decay in the isospin limit. We  therefore turn to  $\gamma^*\to K^+ K^- \pi^0$ instead. 
	The relevant piece reads
	\begin{equation}
		\begin{aligned}
\!\! 			{\cal L}^{(3)} 
			& =
			\frac{  - i     d_s\,\Lambda_2 
			}  { \sqrt2 f_K^2 f_\pi } 
			F^{\mu\nu} 
			\Big[
			K^+\,(\partial_\mu K^-)\,(\partial_\nu\pi^0)
			\Big] + h.c.\,, 
		\end{aligned}
	\end{equation}
	where  
	$ F^{\mu \nu } = \big( \partial ^\mu A^ \nu 
	- \partial ^\nu A^ \mu \big). $
	We focus on this channel because $K^\pm$ is on shell and $\pi^0$ can be reconstructed from two photons in relevant experiments.   
	
	In the SM, the leading contribution is fixed by the
	WZW term. Since this term is the  
	descendant of a five-dimensional Chern--Simons form, it is naturally written in
	the language of differential forms~\cite{Wess:1971yu,Witten:1983tw} 
	\begin{equation}
		\label{eq:WZW-gauged}
		{\cal L}_{\chi{\rm PT}}^{\rm (WZW)}
		=
		\frac{ \,N_c }{48\pi^2}\, 
		{\rm Tr}\!\left[
		e 	Q\,A \wedge   
		\alpha ^3 
		\right]   . 
	\end{equation}
	Here $A$  stands for the photon field. 
	The overall normalization is fixed by the
	underlying chiral anomaly. Equivalently, it is topologically quantized by the
	five-dimensional Chern--Simons construction, so the coefficient is an integer, which is $N_c$ in QCD. 
	Expanding to \({\cal O}(f_P^{-3})\) yields the familiar local
	\(\gamma^\ast\to K^+K^-\pi^0\) vertex
	\begin{equation}
		\label{eq:WZW-KKpi0}
		{\cal L}_{\chi{\rm PT}}^{\gamma KK\pi^0}
		=
		\frac{\sqrt{2}  N_c \,e}{12\pi^2\, f_K^2 f_\pi }\,
		A \wedge  
		d  \pi^0  \wedge  
		d  K^+  \wedge   
		d K^-  . 
	\end{equation}
	The WZW-induced amplitude
	therefore provides the irreducible odd-parity baseline for
	\(\gamma^\ast \to K^+K^-\pi^0\).  
	The interference between the WZW term and the EDM naturally occurs because both induce odd-intrinsic-parity processes. In contrast, ordinary $\chi$PT conserves intrinsic parity and does not contribute. 
	
	In the chiral limit, $\chi$PT yields 
	\begin{equation} 
		\label{eq:FA}
		\!\!\!\!  	F_V (0,0,0)
		= 
		\frac{ \sqrt{2}  N_c\,e}{12\pi^2\, f_K^2 f_\pi },
		~~ 
		F_A (0,0,0) 
		=
		\frac{    -d_s\,\Lambda_2 }{\sqrt2  f_K^2 f_\pi}\,.
	\end{equation}
Ideally, one can test this mechanism via scattering processes such as $K^+K^- \to \pi^0\gamma$. However, to construct a CP-odd observable, one must measure the photon polarization, and the required initial state is difficult to produce experimentally. Hence, we consider the process $\gamma^* \to K^+K^-\pi^0$ with $\sqrt{s} > 2m_K + m_\pi$. Point-by-point extractions of $F_V$ and $F_A$ can be performed experimentally. Here we approximate them  using a vector-meson-dominance (VMD) model~\cite{Sakurai:1960ju} based on the sequential decays $\gamma^* \to K^+K^{*-}(892)+\text{c.c.}$, given by   
	\begin{eqnarray}\label{VMD}
		  \!\!\!	F_{V,A}\Big(s, s_+ , s_- \Big)  &=&
		F _{V,A}(s,0,0) 
	V(s_+ ,s_-) , 
\nonumber		\\
		\!\!\!
	V(s_+ ,s_-)  &=& 
		\frac{1}{2}  \Big(
		\frac{m_{K^*}^{2}}{m_{K^*}^{2}- s_+  - i \Gamma  _{K^*} \sqrt{s_+}}
\\
&& 	~~+ 
		\frac{m_{K^*}^{2}}{m_{K^*}^{2}-  s_- - i \Gamma  _{K^*} \sqrt{s_- } }
		\Big)\nonumber,  
		\end{eqnarray}
	where 
	$F_{V}(s,0,0)$ is a nonperturbative input to be fitted from the experiments.   
	The
	determination of the ratio 
	$F_A(s,0,0)/F_V(s,0,0) $ needs to await lattice input, and here we    take it as  a constant in $\gamma^*\to K^+K^-\pi^0$. 
	
	For $J/\psi \to K^ + K ^ - \pi ^0$, the leading form factor $F_V$ is induced by three-gluon annihilation, and the ratio $F_A/F_V$ is no longer given by 
	Eq.~\eqref{eq:FA}. 
The leading form factor 
	$F_V(s,0,0)$ is extracted  from the experimental branching fraction, while     we adopt the resonant structure 
	\begin{equation}\label{model}
		F_A ( m_{J/\psi}^2,0 ,0 ) =  
		\frac{eQ_c f_{J/\psi } }{m_{J/\psi} }
		\frac{m_A^2}{ m_A^2 -m_{J/\psi}^2 }	\frac{    d_s\,\Lambda_2 }{\sqrt2  f_K^2 f_\pi}  \,,
	\end{equation}
	where $Q_c=2/3$ is the charge of the charm quark, $f_{J/\psi} = ( 410.4\pm1.7) $~MeV 
	is the decay constant of $J/\psi$~\cite{Hatton:2020qhk}, and 
	$m_A$ is the nominal axial-vector meson mass and we vary it from 1~GeV to 2~GeV as an uncertainty of our framework. 
	The dependencies on $s_+$ 
	and $s_-$ can be well described by the VMD in Eq.~\eqref{VMD}. In particular, the BESIII collaboration found that more than $95\%$ of the ${\cal B}(J/\psi \to K^+ K^- \pi^0)$ is made of the sequential decay $ J/\psi \to K ^ + K^{*-} +c.c.$~\cite{BESIII:2019apb}.

	\section{
		Numerical result 
	}\label{sec4}
	
	We propose to probe the EDM-induced interference in
	$e^+e^- \to   K^+K^-\pi^0$. 
	For a sample of $N$ selected signal events, the purely statistical uncertainty is
	\begin{equation}
		\delta A_T\simeq \sqrt{\frac{1-A_T^2}{N}}
		\;\approx\; \frac{1}{\sqrt{N}}
		\qquad (|A_T|\ll 1)\,.
	\end{equation}
	The relation between $A_T$ and $d_s$ is denoted as 
	\begin{equation}
		d_s   =  
		\frac{ -  N_c e }{6 \,\pi^2 \Lambda _2  }\, \frac{A_T}{R(s)}  \,.
	\end{equation}
Here, $R(s)$ is a phase factor that gauges our ability to extract $d_s$ from $A_T$, and it can be computed numerically from Eq.~\eqref{eq:AT-WD} once the form factors are provided.   
Taking $F_A$ and $F_V$ to be saturated by VMD, we have
\begin{equation}
	R(s)
	=r_{V} 
	\frac{
		\displaystyle \int d\Phi\, |\vec p_0|\, |\vec p_+^{\,\ast}|\,
		|V(s_+,s_-)|^2\, \bigl| \vec V_\perp \!\cdot\! \vec A_\perp \bigr|
	}{
		\displaystyle \int d\Phi\, |\vec p_0|\, |\vec p_+^{\,\ast}|\,
		|V(s_+,s_-)|^2\, \vec V_\perp \!\cdot\! \vec V_\perp
	}\, .
	\label{eq:R_gamma_appendix}
\end{equation}
Here, $r_V$ denotes the factor associated with $V \to K^+ K^- \pi^0$, where $V=\gamma^*, J/\psi$,  given by
\begin{eqnarray}
	\! \!\! \!\!  r_{\gamma^* } &\!=\!& 2  \,, \\ 
	\!\!\! \! r_{J/\psi} &\!=\!& 
	\frac{- \sqrt{2} N_c\, e^2 Q_c f_{J/\psi}\, m_A^2}
	{6\,\pi^2\, f_K^2 f_\pi\, m_{J/\psi}\,
		F_V(m_{J/\psi}^2,0,0)\,\bigl(m_A^2-m_{J/\psi}^2\bigr)}  
	. \nonumber
\end{eqnarray} 
We take 
	$ \Lambda_2  = (18.20^{+1.17}_{-0.99})\times 10^{-3} ~\mathrm{GeV}$~\cite{Jiang:2012ir}.  
	
For CMD--3, the analysed sample ($L\simeq 34~\mathrm{pb}^{-1}$)~\cite{Ryzhenenkov:2017lumi,Erofeev:2017kkpi0}
and the measured Born cross section
$\sigma(e^+e^-\!\to K^+K^-\pi^0)\sim 0.05$--$1.1~\mathrm{nb}$ in
$\sqrt{s}=1.4$--$2.0~\mathrm{GeV}$~\cite{Erofeev:2017kkpi0}
imply
$
N_{\rm prod} \sim (1.7\times10^{3}\!-\!2.7\times10^{4})\,.
$ 
 The VMD behavior
in Eq.~\eqref{VMD} 
 gives $R(s)\approx 1.56 $ at $1.6$~GeV. 
Taking 
the accumulated events over $q^2$ with 
 $N_{\textrm{tot}}\sim 10^{3}$--$10^{4}$, we obtain
\begin{equation}
	\delta d_s \sim (3.5 \text{ -- } 11.1 )\times10^{-16}\,e\cdot\mathrm{cm}\,.
\end{equation}
A prospective VEPP--2000 data set of
$L\sim 1~\mathrm{fb}^{-1}$~\cite{Rastigeev:2024tek} would improved the precision by an order.
The overall normalization largely cancels in asymmetries. 

Similar measurements can be performed at BESIII at center-of-mass energies from 2.00 to 3.08 GeV, where the dominant resonance is expected to be the \(K_2^\ast\)~\cite{BESIII:2022wxz}, with an event yield of \(\mathcal{O}(10^5)\). Here we focus on the \(J/\psi\) resonance, which is dominated by the \(K^\ast(892)\) contribution~\cite{BESIII:2019apb}. 
From the branching fraction of ${\cal B}(J/\psi \to K^+ K^-\pi^0)=
	(2.88 \pm 0.01 \pm 0.12)\times 10^{-3}
	$, we find $ 
	F_V ( m_{J/\psi} ^2, 0,0 ) = -3.7 \times 10 ^{-2} $~GeV$^{-3}$ which gives  
$R(s)=0.8 \text{--}4.7 $   from Eq.~\eqref{model}. 
At BESIII, $182{,}972$ candidate events were observed in 2019~\cite{BESIII:2019apb}, and the $J/\psi$ sample size has increased by about $45$ times since 2009~\cite{BESIII:2021cxx}. The expected statistical precision of $A_T$ is around $3.5\times 10^{-4}$, resulting in
	\begin{equation}
		\delta d_s 
		\;\simeq\; ( 4.2  \text{--} 26.9  )   
		\times 	10^{-18}\;
		e\cdot\mathrm{cm}  \,, 
	\end{equation} 
	which can achieve better constraint than the one from 
	$d_\Lambda$ in Eq.~\eqref{eq1} and 
	ready to be tested at BESIII. 
	The uncertainties  of $d_s$ quoted  here mainly comes from the unknown of $F_A$.   
A solid determination can be achieved once $F_A(s,0,0)$ is determined from a nonperturbative method. 
In general, all CP violating operators may contribute to $F_A$ and may affect the constraint for $d_s$ here. 
Other CP-odd operators without explicit photon fields (e.g. chromo-EDMs and four-quark operators) mainly affect purely hadronic observables; here we set them to zero and focus on the  EDM as a minimal benchmark, leaving a systematic EFT treatment to future work.

The Super $\tau$ Charm Facility will increase the production of  $J/\psi $ 
by two 
orders of magnitude~\cite{STCF:2023CDR_PhysDet,STCF:2025CDR_Acc}  and $\delta d_s$  by an order of magnitude, comparing to BESIII. In addition, a border region of $\sqrt s$ range from $4.2$~GeV to $7$~GeV may be explored. 
Complementary information may also come from Belle~II, which aims at an integrated luminosity of $\sim 50~\mathrm{ab}^{-1}$~\cite{Belle2LumiProj2024Dec}, providing enormous statistics for exclusive hadronic final states with excellent charged--hadron identification.  
Although Belle~II operates near the $\Upsilon(4S)$, the initial-state-radiation program enables contiguous measurements of $e^+e^-\to K^+K^-\pi^0$ over a wide effective $\sqrt{s}$ range~\cite{BaBar:2007ceh}. 
	
\section{Conclusion} 
	\label{sec5}	
We develop a framework to probe quark EDMs in $\gamma^\ast \to PPP$ using a gauge-invariant basis of transverse vectors. We identify the CP-odd and T-odd interference observable $\vec V_\perp\!\cdot\!\vec A_\perp$. As a concrete example, we study $\gamma^\ast\to K^+K^-\pi^0$ in $\chi$PT: the vector current is fixed by the WZW term, while the EDM induces an axial current through the chiral realization of $d_s$. We match the quark-level operator onto chiral fields and obtain the form factors in the chiral limit. The resulting interference between the topological anomaly and $d_s$ is expected, since both violate intrinsic parity.

We define a signed interference asymmetry $A_T$ to isolate EDM effects and compute the statistical sensitivity $\delta d_s$. Numerical estimates for $e^+e^-\to K^+K^-\pi^0$ at CMD-3 indicate sensitivities at the $10^{-16}\,e\cdot\mathrm{cm}$ level with current data, and improved reach with higher luminosity. We extend the analysis to $J/\psi\to K^+K^-\pi^0$ and note that existing BESIII samples could reach $\delta d_s\sim 10^{-18}\,e\cdot\mathrm{cm}$. The asymmetry construction cancels the overall normalization and mitigates many detector effects, providing a robust strategy to probe CP-violating strange-quark dipole moments.

	\section*{Acknowledgements}
This work is supported in part by the National Natural Science Foundation of China (NSFC) under Grant Nos. 12547104 and 12575096; and the China Postdoctoral Science Foundation (CPSF) under Grant No. 2025M773361.

\end{document}